\documentclass[UKenglish,runningheads,envcountsect,envcountsame]{llncs}
\usepackage{etex,etoolbox}


\usepackage{ifdraft}
\ifdraft{
\usepackage[nohead,nofoot,
            headsep = 1cm,
            footskip = 1cm,
            text={12.2cm,19.3cm},
            paperwidth=16.2cm,
            paperheight=23.3cm,
            marginparsep=1mm,
            marginparwidth=1.88cm,
            footskip=1cm,
            centering
            ]{geometry}
}{}

\usepackage{hyperref}
\hypersetup{hidelinks,final}

\usepackage{graphicx,nicefrac}
\usepackage[notref,notcite,final]{showkeys}
\usepackage{cite}
\usepackage{mathtools}

\usepackage{seqsplit}
\usepackage{xstring}
\newcommand{\defaultshowkeysformat}[1]{%
\StrSubstitute{#1}{ }{\textvisiblespace}[\TEMP]%
\parbox[t]{\marginparwidth}{\raggedright\normalfont\small\ttfamily\(\{\){\color{red!50!black}\expandafter\seqsplit\expandafter{\TEMP}}\(\}\)}%
}

\newcommand\bsfrac[2]{%
\scalebox{-1}[1]{\nicefrac{\scalebox{-1}[1]{$#1$}}{\scalebox{-1}[1]{$#2$}}}%
}

\renewcommand*\showkeyslabelformat[1]{%
\noexpandarg%
\defaultshowkeysformat{#1}%
}

\newcommand{\osOne}{\gamma}
\newcommand{\osZero}{\Theta}
\newcommand{\prios}{2nk}

\newcommand{\Pow}{\mathcal{P}}

\usepackage[inline]{enumitem}
\setlist[enumerate,1]{label=(\arabic*),font=\normalfont,align=left,leftmargin=0pt,labelindent=0pt,listparindent=\parindent,labelwidth=0pt,itemindent=!,topsep=3pt,parsep=0pt,itemsep=3pt,start=1}
\setlist[enumerate,2]{label=(\alph*),font=\normalfont,labelindent=*,leftmargin=*,start=1}
\setlist[itemize]{labelindent=*,leftmargin=*,topsep=5pt,itemsep=3pt}
\setlist[description]{labelindent=*,leftmargin=*,itemindent=-1 em}

\usepackage{xspace}
\usepackage{xcolor}
\usepackage{soul}
\usepackage[utf8]{inputenc}
\usepackage{amsmath,amssymb}
\usepackage{stmaryrd} 
\usepackage{wasysym}

\numberwithin{equation}{section}

\usepackage[author=anonymous,marginclue,footnote]{fixme}
\FXRegisterAuthor{dh}{adh}{DH}
\FXRegisterAuthor{ls}{als}{LS}
\FXRegisterAuthor{mg}{am}{DH}

\usepackage{algorithmicx}
    \usepackage[ruled]{algorithm}
\usepackage[noend]{algpseudocode}

\usepackage{tikz}
 \usetikzlibrary{trees}
 \usetikzlibrary{shapes}
 \usetikzlibrary{fit}
 \usetikzlibrary{shadows}
 \usetikzlibrary{backgrounds}
 \usetikzlibrary{arrows,automata}

\tikzset{
   n/.style= {circle,fill,inner sep=1.5pt,node distance=2cm}
  ,acc/.style={circle,draw,inner sep=3pt,node distance=2cm}
  ,phantom/.style={circle},
  ,arr/.style={->, >=stealth, semithick, shorten <= 3pt, shorten >= 3pt}
}
\usepackage{pgfplots,pgfplotstable}

\renewcommand{\Box}{\square}
\renewcommand{\Diamond}{\lozenge}

\newcommand{\hearts}{\heartsuit}

\newcommand{\sem}[1]{\llbracket #1 \rrbracket}

\makeatletter
\def\moverlay{\mathpalette\mov@rlay}
\def\mov@rlay#1#2{\leavevmode\vtop{%
   \baselineskip\z@skip \lineskiplimit-\maxdimen
   \ialign{\hfil$\m@th#1##$\hfil\cr#2\crcr}}}
\newcommand{\charfusion}[3][\mathord]{
    #1{\ifx#1\mathop\vphantom{#2}\fi
        \mathpalette\mov@rlay{#2\cr#3}
      }
    \ifx#1\mathop\expandafter\displaylimits\fi}
\makeatother

\newcommand{\detcarrier}{D_\target}
\newcommand{\detprio}{\Omega}

\newcommand{\target}{\chi}

\newcommand{\Bag}{\mathcal{B}}

\newcommand{\Sem}[1]{{[\![#1]\!]}}

\newcommand{\FaCT}[0]{FaCT{\footnotesize++}}

\newcommand*{\ATLquant}[1]{\langle\!\langle#1\rangle\!\rangle}

\newcommand{\takeout}[1]{\empty}

\spnewtheorem{assumptions}[theorem]{Assumptions}{\bfseries}{\rmfamily}
\spnewtheorem{notation}[theorem]{Notation}{\bfseries}{\rmfamily}
\spnewtheorem{observation}[theorem]{Observation}{\bfseries}{\rmfamily}
\spnewtheorem{defn}[theorem]{Definition}{\bfseries}{\rmfamily}
\spnewtheorem{expl}[theorem]{Example}{\bfseries}{\rmfamily}
\spnewtheorem{rem}[theorem]{Remark}{\bfseries}{\rmfamily}
\spnewtheorem{fact}[theorem]{Fact}{\bfseries}{\rmfamily}
\spnewtheorem{construction}[theorem]{Construction}{\bfseries}{\rmfamily}

\newcommand\ExpTime{$\textsc{ExpTime}$\xspace}

\begin{document}

\title{COOL~2 -- A Generic Reasoner\\ for Modal Fixpoint Logics (System Description)}
\titlerunning{COOL~2 System Description}
 \author{%
   Oliver G\"orlitz\inst{1}
   \and Daniel Hausmann \orcidID{0000-0002-0935-8602} \inst{2} 
   \thanks{Supported by  the ERC Consolidator grant D-SynMA (No.
	772459)}\and Merlin Humml \orcidID{0000-0002-2251-8519} \inst{1} \thanks{Supported by Deutsche Forschungsgemeinschaft (DFG) as part of the Research Training Group~2475 (grant number~393541319/GRK2475/1-2019) and the project `RAND' (grant number~377333057).}
   \and Dirk Pattinson \orcidID{0000-0002-5832-6666} \inst{3}
   \and Simon Prucker \orcidID{0009-0000-2317-5565} \inst{1}
   \and Lutz Schr\"oder \orcidID{0000-0002-3146-5906} \inst{1} \thanks{Supported by Deutsche Forschungsgemeinschaft (DFG) under project MI 717/7-1}
 }
\authorrunning{G\"orlitz, Hausmann, Humml, Pattinson, Prucker, Schr\"oder}
\institute{Friedrich-Alexander-Universit\"at Erlangen-N\"urnberg, Germany \and
  Gothenburg University, Sweden \and Australian National University, Canberra, Australia}

\maketitle
\begin{abstract}
  There is a wide range of modal logics whose semantics goes beyond
  relational structures, and instead involves, e.g., probabilities,
  multi-player games, weights, or neighbourhood
  structures. Coalgebraic logic serves as a unifying semantic and
  algorithmic framework for such logics. It provides uniform reasoning
  algorithms that are easily instantiated to particular, concretely
  given logics.  The \emph{COOL~2} reasoner provides an implementation
  of such generic algorithms for coalgebraic modal fixpoint logics.
  As concrete instances, we obtain in particular reasoners for the
  aconjunctive and alternation-free fragments of the graded
  $\mu$-calculus and the alternating-time $\mu$-calculus.  We evaluate
  the tool on standard benchmark sets for fixpoint-free graded modal
  logic and alternating-time temporal logic (ATL), as well as on a dedicated 
  set of benchmarks for the graded
  $\mu$-calculus.
\end{abstract}
%
\noindent
\section{Introduction}

\noindent Modal and temporal logics are established tools in the
specification and verification of systems. While many such
logics 
are interpreted over relational transition systems, the semantics of
quite a number of important logics goes beyond the relational setup,
involving, for instance,
probabilities~\cite{LarsenSkou91,HanssonJonsson94}, concurrent games
as in alternating-time logics~\cite{AlurEA02,Pauly02},
monotone neighbourhoods structures as in game logic~\cite{Parikh83}
and concurrent dynamic logic~\cite{Peleg87}, or integer transition
weights as in the multigraph semantics~\cite{DAgostinoVisser02} of the
graded $\mu$-calculus~\cite{KupfermanEA03}. \emph{Coalgebraic
  logic}~\cite{CirsteaEA11} provides a uniform semantic and
algorithmic framework for these logics, based on the paradigm of
\emph{universal coalgebra}~\cite{Rutten00}. It provides reasoning
algorithms of optimal complexity at various levels of expressiveness,
up to 
the coalgebraic
$\mu$-calculus~\cite{CirsteaEA11a,HausmannEA16,HausmannEA18,HausmannSchroder19}. These
algorithms are parametric in the transition type of systems (weighted,
probabilistic, game-based etc.) as well as in suitable choices of
modalities specific to the given system type. Their instantiation to
specific logics requires providing either a set of next-step modal
tableau rules satisfying a suitable completeness
criterion~\cite{SchroderPattinson09b} or, more generally, a plug-in
algorithm that determines satisfiability for an extremely simple
\emph{one-step logic} that describes the interaction between
modalities, and consists of (conjunctions of) modal operators applied
to variables only~\cite{KupkeEA22}.

The \emph{COalgebraic Ontology Logic solver (COOL)} provides reasoning
support for coalgebraic logics based on these generic algorithms. The
first version of the tool~\cite{GorinEA14} provided reasoning support
for fixpoint-free coalgebraic hybrid logic with global assumptions,
using a global caching principle~\cite{GoreEA10b}. In the present
paper, we present \emph{COOL~2}, which provides reasoning support for
coalgebraic fixpoint logics, specifically for both the aconjunctive
fragment and the alternation-free fragment of the coalgebraic
$\mu$-calculus.  By instantiation, we obtain in particular the first
implemented reasoners for the graded
$\mu$-calculus~\cite{KupfermanEA02} (for which a set of coalgebraic
modal tableau rules has been described in the
literature~\cite{SchroderPattinson09b}; however, this rule set has
later turned out to be incomplete, cf.\ \autoref{rem:graded}) and the
alternating-time $\mu$-calculus~\cite{AlurEA02}. We describe the
structure of the tool including implementational details, and present
evaluation results, focusing on the graded $\mu$-calculus and
alternating-time temporal logic (ATL).  Additional details on the
evaluation can be found in the appendix.

\paragraph{Related Work:}
We have already mentioned work in coalgebraic logic on which COOL is
based~\cite{SchroderPattinson09b,GoreEA10b,CirsteaEA11a,HausmannEA16,HausmannEA18,HausmannSchroder19}. COOL
is conceptually a successor of the \emph{Coalgebraic Logic
  Satisfiability Solver (CoLoSS)}~\cite{CalinEA09} but does not share
any of its code. CoLoSS implements fixpoint-free logics, and is
entirely unoptimised. The first version of COOL~\cite{GorinEA14} has been evaluated on
fixpoint-free next-step logics.

COOL does cover also various relational modal logics, for which there
are numerous specialised reasoners, including highly optimised
description logic reasoners such as \FaCT~\cite{Tsarkov:2006:FDL},
Pellet~\cite{SirinEA07}, RACER~\cite{HaarslevMoeller01}, and
HermiT~\cite{GlimmEA14}. As these systems do not support fixpoint
logics, a comparison would be of limited value.  In previous work,
COOL has been evaluated on various relational fixpoint logics, and has
been shown to perform favourably on Computation Tree Logic~\cite{HausmannEA16} (in comparison to reasoners featured in a
previous systematic evaluation~\cite{GoreEA11}), as well as on the
aconjunctive fragment of the modal $\mu$-calculus~\cite{HausmannEA18}
(in comparison to MLSolver~\cite{FriedmannLange10MLSolver}).  A
reasoner for (next-step) graded modal logic has been evaluated against
various description logic reasoners~\cite{SnellEA12}, using however
the above-mentioned incomplete set of modal tableau rules.

For the same reasons, we refrain from evaluating COOL~2 against reasoners
for coalition logic, i.e.\ the fixpoint-free fragment of the
alternating-time $\mu$-calculus, such as CLProver~\cite{NalonEA14}.
The only implemented reasoner for any fragment of the alternating-time
$\mu$-calculus that does include fixpoints still appears to be the tableau
reasoner TATL for alternating-time temporal logic~\cite{David13,David15}. 
TATL has been compared to COOL on random formulas in previous work~\cite{HausmannEA16}.




\section{Satisfiability in the Coalgebraic $\mu$-Calculus}\label{sec:mu}

COOL~2 is a satisfiability checker for the coalgebraic
$\mu$-calculus~\cite{CirsteaEA11a}, that is, for the extension of
coalgebraic modal logic with extremal fixpoint operators. Formulas of
this logic are interpreted over coalgebras, where the semantics of
modal operators is defined by means of so-called \emph{predicate
  liftings}~\cite{SchroderPattinson09b}; we recapitulate examples of
system types and modalities subsumed by this paradigm in \autoref{ex:logics}. 

\paragraph{Syntax:} Formulas are built relative to a set
$\mathsf{Var}$ of fixpoint variables and a \emph{modal similarity
  type} $\Lambda$, that is, a set of modal operators with assigned
finite arities that is closed under duals, with
$\overline{\hearts}\in\Lambda$ denoting the dual of
$\hearts\in\Lambda$.  Formulas $\psi,\phi,\dots$ of the
\emph{coalgebraic $\mu$-calculus} over~$\Lambda$ are given by the
grammar
\begin{align*}
\psi,\phi:= \bot\mid\top\mid \psi\land\phi\mid\psi\lor\phi\mid\hearts(\psi_1,\dots,\psi_n)\mid X\mid \mu X.\,\psi\mid
\nu X.\,\psi,
\end{align*}
where $\hearts\in\Lambda$ has arity~$n$ and $X\in\mathsf{Var}$. A formula
$\target$ is \emph{aconjunctive} if for every conjunction
$\psi\land\phi$ that is a subformula of~$\target$, at most one of the
formulas~$\psi$ and~$\phi$ contains a free fixpoint variable $X$ that is bound
by a least fixpoint operator $\mu X.\,$ 
While the logic does not contain negation as
an explicit operator, full negation can be defined as usual; e.g. we
have $\neg\hearts\psi=\overline{\hearts}\neg\psi$ and
$\neg\mu X.\,\psi = \nu X.\,\neg\psi[\neg X/X]$, using $\neg\neg X=X$.

Both the theoretical satisfiability checking algorithm and its
implementation in COOL~2 operate on the \emph{Fischer-Ladner
  closure}~\cite{Kozen83,HausmannSchroder19,KupkeEA22b} of the target
formula.  The \emph{alternation depth}
(e.g.~\cite{Niwinski86,HausmannSchroder19,KupkeEA22}) of a formula is
the maximum depth of dependent alternating nestings of least and
greatest fixpoints within the formula. Formulas with alternation
depth~$1$ are \emph{alternation-free}.

\paragraph{Semantics:} Formulas are interpreted over $F$-coalgebras, that is, structures 
\begin{align*}
(C,\xi:C\to FC),
\end{align*} 
where $F\colon \mathsf{Set}\to\mathsf{Set}$ is a functor determining
the branching type of the systems at hand; thus $\xi(x)\in FC$ encodes
the transitions from $x\in C$, structured according to~$F$. Modalities
$\hearts\in\Lambda$ of arity~$n$ are interpreted as \emph{predicate liftings}, that
is, families of maps $\sem{\hearts}_U\colon (2^U)^n\to 2^{FU}$ (for
$U\in\mathsf{Set}$) that assign predicates on~$FU$ to $n$-tuples of predicates on~$U$, subject to a \emph{naturality}
condition~\cite{Pattinson04,Schroder08}.  On a coalgebra $(C,\xi)$,
the semantics of formulas is defined 
inductively in the usual
way for the propositional operators and fixpoints, and by 
$\sem{\hearts(\psi_1,\dots,\psi_n)}=\xi^{-1}[\sem{\hearts}_C(\sem{\psi_1},\dots,\sem{\psi_n})]$
for modalities.

A closed formula $\psi$ is \emph{satisfiable} if there is a coalgebra $(C,\xi)$ and a state $x\in C$
such that $x\in\sem{\psi}$. A formula $\psi$ is \emph{valid} if  $\neg\psi$ is not
satisfiable.

\begin{expl}\label{ex:logics}
\begin{enumerate}
\item The standard \emph{modal $\mu$-calculus}~\cite{Kozen83} is obtained using the functor $F=\mathcal{P}(A)\times \mathcal{P}$,
where $A$ is a fixed set of atoms, 
the similarity type $\Lambda=\{\Diamond,\Box,a,\neg a\mid a\in A\}$, and predicate liftings%
\begin{small}
\begin{align*}
\sem{\Diamond}_C(B) &= \{(A,Z)\in 2^A\times 2^C\mid Z\cap B\neq\emptyset\} &
\sem{a}_C & = \{(A,Z)\in 2^A\times 2^C\mid a\in A\} \\
\sem{\Box}_C(B) &= \{(A,Z)\in 2^A\times 2^C\mid Z\subseteq B\} &
\sem{\neg a}_C & = \{(A,Z)\in 2^A\times 2^C\mid a\notin A\}
\end{align*}
\end{small}%
The expressive power of the modal $\mu$-calculus is demonstrated by the formulas
\begin{align*}
&\mu X.\,\nu Y.\, (p \land \Diamond Y) \lor \Diamond X &
&\nu X.\,\mu Y.\, (p \land \Diamond X) \lor \Diamond Y.
\end{align*}
The former is a co-B\"uchi formula expressing the existence of a path on which~$p$ holds 
forever, from some point on; the latter formula  expresses the B\"uchi property that there
is a path on which the atom $p$ is satisfied infinitely often.

\item The \emph{graded $\mu$-calculus}~\cite{KupfermanEA02} allows
  expressing quantitative properties with the help of modal operators
  $\langle n \rangle$ and $[n]$, $n\in\mathbb{N}$; formulas
  $\langle n\rangle\psi$ and $[n]\psi$ then have the intuitive meaning
  that `there are more than $n$ successor states that satisfy~$\psi$',
  and `all but at most $n$ successor states satisfy~$\psi$',
  respectively. Its coalgebraic interpretation is based on
  \emph{multigraphs}, which are coalgebras for the multiset
  functor~\cite{DAgostinoVisser02}. A graded variant of the above
  B\"uchi property is specified, e.g., by the formula
  $\nu X.\,\mu Y.\, (p \land \langle n\rangle X) \lor \langle n\rangle
  Y$, which expresses the existence of an infinite $n+1$-ary tree such
  that the atom $p$ is satisfied infinitely often on every path in the
  tree.

\item The \emph{alternating-time $\mu$-calculus} (AMC)~\cite{ScheweThesis} 
extends coalition logic~\cite{Pauly02} with fixpoints
and (modulo syntax) supports modalities $\langle D\rangle$ and $[D]$, where $D\subseteq N$ is a coalition formed
by agents from the set $N=\{1,\ldots,n\}$ for some fixed $n\in\mathbb{N}$; formulas
$\langle D\rangle\psi$ and $[D]\psi$ then state that `coalition $D$ has a joint strategy to enforce $\psi$'
and that `coalition $D$ cannot prevent $\psi$', respectively. 
For instance, the formula
$\nu X.\,\mu Y.\,\nu Z.\, (p \land \langle D\rangle X) \lor (q\wedge \langle D\rangle Y)\lor (\neg q \land \langle D\rangle Z)$
expresses that coalition $D$ has a joint multi-step strategy that guarantees that $p$ is visited infinitely often whenever
$q$ is visited infinitely often.
\end{enumerate}
\end{expl}

\paragraph{Satisfiability Checking:}\label{par:sat}

We proceed to recall the satisfiability checking algorithm for the
coalgebraic $\mu$-calculus that forms the basis of the implementation
within COOL~2.  This algorithm adapts the automata-based approach to
satisfiability checking for the standard $\mu$-calculus, and
generalises the treatment of modal steps by parametrizing over a
solver for the \emph{one-step satisfiability} problem of the logic,
which concerns satisfiability of formulae with exactly one layer of
next-step modalities~\cite{HausmannSchroder19}. It thus avoids the
necessity of tractable sets of tableaux rules for modal
operators. Under mild assumptions on the complexity of the one-step
satisfiability problem of the base logic at
hand (`\emph{tractability}'), the algorithm witnesses a, typically
optimal, upper bound \ExpTime for the complexity of the satisfiability
problem; unlike a previous algorithm~\cite{CirsteaEA11}, the algorithm
thus has optimal runtime also in cases where no tractable sets of
modal tableaux rules are known, such as the graded (or, more
generally, Presburger) $\mu$-calculus (further cases of this kind
include the probabilistic $\mu$-calculus with polynomial
inequalities~\cite{HausmannSchroder19} and the unrestricted form of
the \emph{alternating-time $\mu$-calculus with disjunctive explicit
  strategies}~\cite{GottlingerEA21}).

The algorithm constructs and solves a parity game
 that characterises satisfiability of the input formula $\target$. 
In this game one player attempts
to construct a tableau structure for $\target$ while the opposing
player attempts to refute the existence of such a structure. Modal steps
in this tableau construction are treated by using instances of the one-step satisfiability problem
for the logic at hand, thereby generalising traditional modal tableau rules.
The winning condition of the game is encoded by a non-deterministic parity automaton
$\mathsf{A}_\target$, reading infinite words that encode
sequences of step-wise formula evaluations (so-called \emph{formula traces})
within a coalgebra; such words encode branches in the constructed tableau structure.
Conjunctions give rise to
nondeterminism in this automaton, and the parity condition of the automaton is used to accept
exactly those words that encode sequences of formula evaluations in which some
least fixpoint is unfolded infinitely often.
To use the language accepted by $\mathsf{A}_\target$ as the winning condition in a parity game, 
we transform $\mathsf{A}_\target$ to an equivalent 
deterministic parity automaton $\mathsf{B}_\target$.
This automaton then is paired with the tableau construction
to yield a parity game in which
the existential player aims to show
the existence of a tableau structure in which all branches are rejected by $\mathsf{B}_\target$, and that is built in such a
way that modalities always are jointly one-step satisfiable. To ensure the latter property,
the modal moves in the game invoke instances of the one-step satisfiability problem of the base logic. 
For more details on one-step satisfiability and the overall algorithm, see the appendix as well as~\cite{HausmannSchroder19}.
%

\begin{corollary}[\hspace{-0.5pt}\cite{HausmannSchroder19}]
Suppose that the one-step satisfiability problem is  tractable. Then the satisfiability problem of
the corresponding instance of the coalgebraic $\mu$-calculus is in \ExpTime.
\end{corollary}


\begin{rem}\label{rem:rules}  \label{rem:graded}
  As mentioned above, previous algorithms for the coalgebraic
  $\mu$-calculus (also implemented in COOL~2) rely on complete sets of
  modal tableau rules, specifically on one-step cutfree complete sets
  of so-called \emph{one-step rules}~\cite{SchroderPattinson09b}; such
  rules (in their incarnation as tableau rules) have a premiss with
  exactly one layer of modal operators and a purely propositional
  conclusion. A typical example is the usual tableau rule for the
  modal logic~$K$: `To satisfy
  $\Box a_1\land\dots\land\Box a_n\land\neg\Box a_0$, satisfy
  $a_1\land\dots\land a_n\land\neg a_0$'. It has been shown that the
  existence of a tractable one-step cutfree complete set of one-step
  rules implies tractability of one-step
  satisfiability~\cite{KupkeEA22}, i.e.~the approach via one-step
  satisfiability is more general.

  As indicated in the introduction, a tractable one-step cutfree
  complete set of one-step rules for graded modal logic has been
  claimed in the literature~\cite{SchroderPattinson09b,SnellEA12} but
  has since turned out to be incomplete; we give a counterexample in
  the appendix. (A similar rule for Presburger
  modal logic~\cite{KupkePattinson10} has also been shown to be in
  fact incomplete~\cite{KupkeEA22}.)
\end{rem}

\section{Implementation}

The previous version COOL~\cite{GorinEA14} only implements fixpoint-free
(coalgebraic) logics, such as standard modal logic, probabilistic modal logic, or
coalition logic. The main novelty of the new version COOL~2,
described here, is 
\begin{itemize}
\item the addition of fixpoint constructs to the previously implemented logics,
supporting alternation-free and aconjunctive fragments of the resulting $\mu$-calculi, and implementing  on-the-fly solving to allow early termination
\item support for treating modal steps both by tableaux rules (when a suitable rule set exists),
and by one-step satisfiability checking (in the remaining cases)
\end{itemize}
In more detail, COOL~2 is written in OCaml and implements the satisfiability checking algorithm described in \autoref{par:sat},
treating modal steps by solving instances of the one-step satisfiability problem\footnote{Sources are available at \url{https://git8.cs.fau.de/software/cool}}. For logics where a suitable set of modal
tableau rules is implemented, those are used for the treatment of modal steps, rather than relying on one-step satisfiability (unless the user explicitly chooses otherwise); in these cases, COOL~2 essentially implements the algorithm described in~\cite{KupkeEA22}.
The current implementation supports the alternation-free and the aconjunctive fragments of the standard $\mu$-calculus (both serial and non-serial), the monotone $\mu$-calculus~\cite{HansenEA17}, the alternating-time $\mu$-calculus (i.e. coalition logic with fixpoint operators), and the graded $\mu$-calculus.
Tractable tableaux rules are available for all cases except for the graded $\mu$-calculus, for which COOL~2 uses
the one-step satisfiability algorithm to decide satisfiability. In particular, COOL~2 is the
only existing reasoner for the graded $\mu$-calculus (as well as the only reasoner covering the alternating-time $\mu$-calculus beyond ATL).

The concrete logic used can be selected via a command-line parameter setting up the data
structures in COOL~2 accordingly before parsing and checking the syntax of the given formula~$\target$.
COOL~2 then builds the determinised automaton~$\mathsf{B}_\target$, yielding
the parity game described above in a step-wise manner, repeatedly adding nodes in \emph{expansion steps} that explore the game.
In the case of simpler alternation-free formulas, the Miyano-Hayashi method~\cite{MiyanoHayashi1984} is used to construct~$\mathsf{B}_\target$, resulting in asymptotically smaller games with a B\"uchi winning condition; for the
more involved aconjunctive formulas, the implementation uses the permutation method for determinisation of limit-deterministic parity
automata~\cite{EsparzaKRS22,HausmannEA18}.
Nodes in the
constructed game are marked as either unexpanded, undecided, unsatisfiable, or satisfiable. 

Optional \emph{solving steps} may take place at any point during the construction of $\mathsf{B}_\target$, depending on runtime parameters of COOL~2; these steps compute the winning regions of the partial game that has been constructed so far and accordingly mark nodes as satisfiable or unsatisfiable, if possible. The reasoner terminates as soon as the initial node is marked satisfiable or
unsatisfiable. If this does not allow for early termination, the game eventually becomes fully explored, at which point a final (obligatory) solving step for the complete game
is guaranteed to mark the initial node, thereby ensuring termination.

We detail the implementation of the two main procedures within COOL~2.

\paragraph{Implementation of Expansion Steps.}
The propositional expansion steps in the game construction for nodes $v$ are
performed using the propositional satisfiability solver
$\mathsf{MiniSat}$~\cite{EenSoerensson03} to compute a word that encodes consistent propositional
formula manipulations for $v$.
Afterwards, the successor of $v$ in $\mathsf{B}_\target$ under this word
is computed and added to the game.

When the one-step satisfiability based algorithm of COOL~2 is used, modal expansion steps
for nodes $v$ create new game nodes for each subset $\kappa$ of the modalities that are
to be jointly satisfied at $v$; this is done by computing the successor of $v$ in $\mathsf{B}_\target$ that is reached by manipulating each formula from $\kappa$.

When the tableau-based algorithm of COOL~2 is used, the modal expansion step for a node $v$ 
instead computes all applications of a modal rule matching $v$ and inserts, for each such rule
application, and each conjunctive clause $\kappa$ in the conclusion of the rule application,
the new game node that is reached from~$v$ in~$\mathsf{B}_\target$ by manipulating the
modalities that constitute~$\kappa$.
Intuitively, using tableau rules reduces the search space by only adding nodes found in the conclusion of some matching rule application.

Any node that is added by some expansion step is initially marked as undecided.
Crucially, all expansion
steps perform on-the-fly determinisation, that is, given a game node $v$ and
a word that encodes a sequence of formula manipulations, the newly added game node is computed using only 
the information stored in $v$. 

\paragraph*{Implementation of Solving Steps.}
A single solving step computes the winning regions in the
parity game that has been constructed up to this point, and marks nodes
accordingly. The game
solving is done using either the parity game solver $\mathsf{PGSolver}$~\cite{FriedmannLange09}
or a native implementation provided by COOL~2 that solves the game by fixpoint iteration.

If the one-step satisfiability-based  algorithm is used,
an assigned modal node $v$ is satisfiable
if its modalities are jointly one-step satisfiable in those successors of $v$ that
are satisfiable themselves.
An enumerative representation of the game thus contains existential moves to
all subsets $\Pi$ of subsets of modalities of $v$ that are sufficiently large for
one-step satisfaction of the modalities of $v$, followed by universal moves to nodes induced by any $\kappa\in \Pi$; the full
game thus is of doubly-exponential size. This can be avoided by inlining the modal steps,
thereby evading the intermediate nodes $\Pi$. The winning region can then be computed in
single-exponential time by using COOL~2's native fixpoint iteration over a function
that computes the two-tiered modal steps in one go.

Decision procedures
for the one-step satisfiability problems in the relational and the graded case are implemented
in COOL~2 along the lines of the algorithms described in~\cite[Example 6]{HausmannSchroder19} (in the graded case, nondeterministic guessing is replaced with a recursive search procedure).

If the algorithm based on modal tableau rules is used, the treatment of modal steps
follows the tableaux-based algorithm that is given in~\cite{CirsteaEA11a}. States $v$ are
satisfiable if for all rule applications that match $v$, the conclusion of the application
contains a conjunctive clause $\kappa$ such that the node induced by $\kappa$ 
is satisfiable.

COOL~2 also allows the user to specify the desired frequency of optional game solving steps,
including the options $\mathsf{once}$ and $\mathsf{adaptive}$. 
With the option $\mathsf{once}$, no intermediate solving takes place so that
the game is fully constructed and solved just once, at the
very end of the execution. 
With the option $\mathsf{adaptive}$, intermediate solving takes places, but the frequency of solving reduces as the size of the constructed graph increases; this option implements \emph{on-the-fly} solving and allows for finishing
early in cases where a small model or refutation exists.

\section{Evaluation}\label{sec:eval}

We conduct experiments in order to evaluate
the performance of the various algorithms implemented in COOL
in comparison with each other, as well as in comparison 
with other tools (where applicable).\footnote{Scripts and executables that allow for reproducing our experiments can be found at \href{https://doi.org/10.5281/zenodo.8042581}{DOI 10.5281/zenodo.8042581}} Complete definitions of all formula series used in the evaluation as well as additional experimental results can be found in the appendix.
\paragraph{Experiments:}
In a first experiment, we compare COOL~2 with the established 
reasoner \FaCT, which supports the description logic $\mathcal{SROIQ}(\mathcal{D})$ 
(subsuming fixpoint-free graded modal logic), using the following series of formulas from Snell~et~al.\cite{SnellEA12}.
{\small
\begin{align*}
  \mathsf{Cardinality}(n) &:= \langle n - 1 \rangle\neg p \land \langle n - 1 \rangle p \land [n]\neg q \land [ n ] q \tag{Sat}\\
  \mathsf{CardinalityU}(n) &:= \langle n - 1 \rangle\neg p \land \langle n - 1 \rangle p \land [n]\neg q \land [ n - 1] q \tag{UnSat}
\end{align*}
}%
\noindent Intuitively, the satisfiable $\mathsf{Cardinality}(n)$ formulas
express that there are at least $2n$ successors and that both $q$ and
$\neg q$ are satisfied in at most $n$ successors, each; similarly the
unsatisfiable $\mathsf{CardinalityU}(n)$ formulas state that there are at
least $2n$ successors, and that $q$ and $\neg q$ hold in at most $n$ and $n-1$ successors,
respectively; the latter statements imply that there are 
at most $2n-1$ successors, yielding a contradiction.
%

Going beyond next-step formulas, we continue by devising
various complex series of graded $\mu$-calculus formulas that involve (nested) fixpoints 
and express non-trivial properties of graded trees, automata and games. 
\begin{itemize}[wide]
\item
We obtain a series of unsatisfiable formulas by requiring the existence of an $n+1$-branching tree in which $p$ holds everywhere while at the same time requiring that this tree  contains some state with $n+2$ successors that satisfy $p$:
{\small
\begin{align*}
\mathsf{TreeU}(n)&=
(\nu X.\, \langle n \rangle (p\wedge X)\wedge[n+1]\neg p) \land (\mu Y.\,\langle n+1\rangle p\vee\langle
n\rangle (p\wedge Y))\tag{UnSat}
\end{align*}
}
\item 
Next we turn our attention to graded formulas involving parity conditions.
We devise a series of valid formulas expressing that graded parity automata can be transformed 
to graded B\"uchi automata
accepting a superlanguage of the original automaton:
{\small
\begin{align*}
\mathsf{ParityToBuechi}(n,k):=\mathsf{Parity}(n,k)\to
\mathsf{Buechi}(n,k)\tag{Valid}
\end{align*}
}%
Here, $\mathsf{Parity}(n,k)$ encodes parity acceptance with $k$ priorities and
grade $n$ while
$\mathsf{Buechi}(n,k)$ expresses B\"uchi acceptance by a nondeterministic
automaton that eventually guesses the maximal priority that
occurs infinitely often; the negated formula
$\neg\mathsf{ParityToBuechi}(n,k)$ is unsafisfiable.

\item Rabin conditions are given by families of pairs $\langle i_j,f_j\rangle_{j\leq k}$ of sets $i_j,f_j$ of states, and express the constraint that there is some $j\leq k$ such that states from $i_j$ (\emph{infinite}) are visited infinitely often and states from $f_j$ (\emph{finite}) are visited only finitely often. We 
can express Rabin conditions with $k$ pairs (and one-step property $\psi$),
B\"uchi properties and satisfaction of single Rabin-pairs
by formulas $\mathsf{Rabin}(k, \psi)$, $\mathsf{Buechi}(f,\psi)$ and $\mathsf{RabinPair}(i,f,\psi)$,
respectively.
 Then we obtain valid formulas stating that the existence of an $n+1$-branching
 tree that satisfies the Rabin condition on each path implies that there is a path
 satisfying a simpler B\"uchi condition or a single Rabin-pair, respectively:
{\small
  \begin{align*}
 \mathsf{RabinToBuechi}(k,n)&:=\mathsf{Rabin}(k,\langle n\rangle)\to\mathsf{Buechi}(i_1\vee\ldots\vee i_k,\langle 0\rangle) \tag{Valid}\\
 \mathsf{RabinToRPair}(k,n)&:= \mathsf{Rabin}(k,\langle n\rangle)\to
 \textstyle\bigvee_{1\leq j\leq k}\mathsf{RabinPair}(i_j,f_j,\langle 0\rangle)\tag{Valid}
 \end{align*}
 }%
\item 
 Coming to games, we specify the winning regions in graded B\"uchi and Rabin games 
 by formulas $\mathsf{BuechiG}(f,n)$ and $\mathsf{RabinG}(k,n)$, respectively;
 in such graded games, players are required to have at least $n$ winning moves at
 their nodes in order to win.
 The following valid formulas then express that winning strategies
in graded Rabin games with $k$ pairs guarantee that some node from $i_1\cup\ldots\cup i_k$ is visited infinitely often:
 {\small
\begin{align*}
 \mathsf{RabinGame}(k,n):=\mathsf{RabinG}(k,n)\to\mathsf{BuechiG}(i_1\vee\ldots\vee i_k,n)\tag{Valid}
 \end{align*}
 }%
\end{itemize}%

In a final experiment on alternating-time formulas, we compare COOL~2 with TATL~\cite{David13} on the ATL example formulas given in~\cite{David13} as well as on additional
formula series.
For instance, we turn the formula $\ATLquant{1}G p \wedge \neg\ATLquant{2}F \ATLquant{1}G p$ (written here using ATL syntax) from~\cite{David13} into a series $\mathsf{Nest}(n)$ with increasing number of nested operators; formulas then alternatingly are satisfiable and unsatisfiable:
\begin{align*}
  \chi(0)&=p &
  \chi(i+1)&=\neg \ATLquant{2} F \ATLquant{1} G \chi(i) &
  \mathsf{Nest}(n) &= \ATLquant{1} G p \land \chi(n),
\end{align*}

\paragraph{Results:} We conducted all experiments on a virtual machine with four \(2,3\)GHz vCPUs processors and 8GB of RAM.
We compare with a 64-bit binary of \FaCT~v1.6.5 and with TATL.
We compute all results with a timeout of $60$ seconds and average the results over multiple executions. 
For the execution and measurement we use hyperfine\footnote{\url{https://github.com/sharkdp/hyperfine}}. 
Below, `COOL' and `COOL on-the-fly' refer to invoking COOL~2 with solving rate $\mathsf{once}$ and $\mathsf{adaptive}$, respectively.
\begin{figure}
\begin{minipage}[t][][t]{.5\linewidth}
  \centering
  \begin{tikzpicture}
    \begin{semilogyaxis}[
      minor tick num=1,
      xtick distance = 4,
      log basis y = 10,
      log ticks with fixed point,
      every axis y label/.style=
      {at={(ticklabel cs:0.5)},rotate=90,anchor=center},
      every axis x label/.style=
      {at={(ticklabel cs:0.5)},anchor=center},
      tiny,
      width=\linewidth,
      height=4.25cm,
      legend columns = 3,
      legend style={at={(0.5,-0.2)},anchor=north},
      ymode=log,
      xlabel={value of $n$},
      ylabel={runtime (s)},
      xmin=1,
      xmax=32,
      ymin=0,
      ymax=60,
      legend entries={COOL on-the-fly,COOL, FaCT++}]

      \addplot [color=blue, mark=*] table [col sep=comma, x=parameter_depth,y=mean] {benchmarks/CARDI-cool-adaptive.csv};
      \addplot [color=red, mark=square*] table [col sep=comma, x=parameter_depth,y=mean] {benchmarks/CARDI-cool-once.csv};
      \addplot [color=black, mark=triangle*] table [col sep=comma, x=parameter_depth,y=mean] {benchmarks/CARDI-fact.csv};
    \end{semilogyaxis}
  \end{tikzpicture}
  \caption{Runtimes for $\mathsf{Cardinality}(n)$}\label{fig:cardi}
\end{minipage}%
\begin{minipage}[t][][t]{.5\linewidth}
  \centering
  \begin{tikzpicture}
    \begin{semilogyaxis}[
      minor tick num=1,
      xtick distance = 2,
      log basis y = 10,
      log ticks with fixed point,
      every axis y label/.style=
      {at={(ticklabel cs:0.5)},rotate=90,anchor=center},
      every axis x label/.style=
      {at={(ticklabel cs:0.5)},anchor=center},
      tiny,
      width=\linewidth,
      height=4.25cm,
      legend columns=3,
      legend style={at={(0.5,-0.2)},anchor=north},
      ymode=log,
      xlabel={value of $n$},
      ylabel={runtime (s)},
      xmin=1,
      xmax=22,
      ymin=0,
      ymax=60,
      legend entries={COOL on-the-fly,COOL, FaCT++}]

      \addplot [color=blue, mark=*] table [col sep=comma, x=parameter_depth,y=mean] {benchmarks/UNCARDI-cool-adaptive.csv};
       \addplot [color=red, mark=square*] table [col sep=comma, x=parameter_depth,y=mean] {benchmarks/UNCARDI-cool-once.csv};
      \addplot [color=black, mark=triangle*] table [col sep=comma, x=parameter_depth,y=mean] {benchmarks/UNCARDI-fact.csv};
    \end{semilogyaxis}
  \end{tikzpicture}
  \caption{Runtimes for $\mathsf{CardinalityU}(n)$}\label{fig:uncardi}
\end{minipage}

\end{figure}

Results for the $\mathsf{Cardinality}$ and $\mathsf{CardinalityU}$ series are shown in \autoref{fig:cardi} and \autoref{fig:uncardi}, respectively. 
From \(n=10\) and \(n=8\) onwards, COOL~2 outperforms \FaCT{} considerably. 
An explanation for this could be that \FaCT{} appears to treat multiplicities in a na\"ive way while COOL~2 employs the more efficient one-step satisfiability algorithm.
\begin{figure}[h]
 \begin{minipage}[t][][t]{.5\linewidth}
 \centering
  \begin{tikzpicture}
    \begin{loglogaxis}[
      log basis y = 10,
      log ticks with fixed point,
      log basis x = 2,
      every axis y label/.style=
      {at={(ticklabel cs:0.5)},rotate=90,anchor=center},
      every axis x label/.style=
      {at={(ticklabel cs:0.5)},anchor=center},
      tiny,
      width=\linewidth,
      height=4.2cm,
      legend columns=2,
      legend style={at={(0.5,-0.22)},anchor=north},
      xlabel={value of $n$},
      ylabel={runtime (s)},
      xmin=2,
      xmax=2^15,
      ymin=0,
      ymax=60,
      xtick distance = 8,
      legend entries={COOL on-the-fly,COOL}]

      \addplot table [col sep=comma, x expr = {2^\thisrow{parameter_size}},y=mean] {benchmarks/TREE-adaptive.csv};
      \addplot table [col sep=comma, x expr = {2^\thisrow{parameter_size}},y=mean] {benchmarks/TREE-once.csv};
    \end{loglogaxis}
  \end{tikzpicture}
  \caption{Runtimes for $\mathsf{TreeU}(n)$}\label{fig:tree}
 \end{minipage}%
 \begin{minipage}[t][][t]{.5\linewidth}
  \centering
  \begin{tikzpicture}
    \begin{loglogaxis}[
      xtick distance = 8,
      log basis y = 10,
      log basis x = 2,
      log ticks with fixed point,
      every axis y label/.style=
      {at={(ticklabel cs:0.5)},rotate=90,anchor=center},
      every axis x label/.style=
      {at={(ticklabel cs:0.5)},anchor=center},
      tiny,
      width=\linewidth,
      height=4.2cm,
      legend columns=3,
      legend style={at={(0.5,-0.2)},anchor=north},
      ymode=log,
      xlabel={$n$ (grades of the modalities)},
      ylabel={runtime (s)},
      xmin=2,
      xmax=2^17,
      ymin=0,
      ymax=60]

      \addlegendentry{$k=1$}
      \addlegendentry{$k=2$}
      \addlegendentry{$k=3$}
      \addlegendentry{$k=4$}

      
      \addplot [color=blue, mark=*] table [col sep=comma, x expr={2^\thisrow{parameter_size}}, y=mean] {benchmarks/PARITYNEG1-adaptive.csv};


      \addplot [color=blue, mark=o] table [col sep=comma, x expr={2^\thisrow{parameter_size}}, y=mean] {benchmarks/PARITYNEG2-adaptive.csv};


      \addplot [color=blue, mark=oplus] table [col sep=comma, x expr={2^\thisrow{parameter_size}}, y=mean] {benchmarks/PARITYNEG3-adaptive.csv};


      
    \end{loglogaxis}
  \end{tikzpicture}
  \caption{Runtimes for $\neg \mathsf{ParityToBuechi}(n,k)$}\label{fig:parity}
\end{minipage}%
\end{figure}

Results for the unsatisfiable tree property are shown in \autoref{fig:tree}. 
As these formulas contain fixpoint operators, a comparison with \FaCT{} is not possible. 
While COOL~2 is generally capable of handling quite large branching factors,
this experiment showcases the drawbacks of on-the-fly solving in the case that a formula cannot be decided early so that repeated attempts of solving the game early lead to overhead computations.

Runtimes for COOL~2 (using on-the-fly solving) on the unsatisfiable series of parity formulas $\neg\mathsf{ParityToBuechi}(n,k)$ are shown in~\autoref{fig:parity}.
The results indicate that
increasing the number of priorities $k$ has a much stronger effect on the runtime than
increasing multiplicities $n$ in the modalities. This is in accordance with expectations
as increasing $k$ leads to much larger determinized automata and resulting satisfiabilty games, while
increasing $n$ only complicates the modal steps in the game while leaving the
global game structure unchanged.

Results for the Rabin families of formulas are given in the table below, with $\dagger$ indicating a timeout of 60 seconds. COOL~2 is able to handle reasonably large formulas describing Rabin properties of automata and games, with the series for $n=1$ expressing properties of standard automata (solved using tableau rules), and the series with $n=2$ properties of graded automata with multiplicity
$2$ (solved using one-step satisfiability).
\begin{figure}
\begin{minipage}[b][][t]{.45\linewidth}
  \centering
\begin{footnotesize}
\begin{tabular}{| l | c | c | c|}
\hline
$\qquad\bsfrac{k}{\text{series}}$ & $1$ & $2$ & $3$\\
\hline
\hline
$\mathsf{RabinToBuechi}(k,1)$ & $0.03$ & $\,\,0.51$ & $45.25$ \\
\hline
$\mathsf{RabinToBuechi}(k,2)$ & $0.08$ & $\,\,10.56$ & $\dagger$\\ 
\hline
\hline
$\mathsf{RabinToRPair}(k,1)$ & $0.03$ & $\,\,8.38$ & $\dagger$\\
\hline
$\mathsf{RabinToRPair}(k,2)$ & $0.07$ & $\dagger$ & $\dagger$\\
\hline
\hline
$\mathsf{RabinGame}(k,1)$ & $0.05$ & $\,\,1.04$ & $\dagger$ \\
\hline
$\mathsf{RabinGame}(k,2)$ & $0.31$ & $43.94$ & $\dagger$ \\
\hline
\end{tabular}
\end{footnotesize}
\end{minipage}
\begin{minipage}[t][][t]{.5\linewidth}
  \centering
  \begin{tikzpicture}
    \begin{semilogyaxis}[
      minor tick num=0,
      xtick distance = 1,
      log basis y = 10,
      log ticks with fixed point,
      every axis y label/.style=
      {at={(ticklabel cs:0.5)},rotate=90,anchor=center},
      every axis x label/.style=
      {at={(ticklabel cs:0.5)},anchor=center},
      tiny,
      width=\linewidth,
      height=3.7cm,
      legend columns=3,
      legend style={at={(0.5,-0.2)},anchor=north},
      ymode=log,
      xlabel={$n$ (nesting depth)},
      ylabel={runtime (s)},
      xmin=0,
      xmax=11,
      ymin=0,
      ymax=60,
      legend entries={COOL on-the-fly,COOL,TATL}]

      \addplot [color=blue, mark=*] table [col sep=comma, x expr=\coordindex, y=mean] {benchmarks/nestedGFCOOL.csv};
      \addplot [color=red, mark=square*] table [col sep=comma, x expr=\coordindex, y=mean] {benchmarks/nestedGFCOOLOnce.csv};
      \addplot [color=black, mark=triangle*] table [col sep=comma, x expr=\coordindex, y=mean] {benchmarks/nestedGFTATL.csv};
            
    \end{semilogyaxis}
  \end{tikzpicture}
  \caption{Runtimes for the ATL series $\mathsf{Nest}(n)$}\label{fig:nestedGF}
\end{minipage}%

\end{figure}

In accordance with previous experiments on random ATL formulas of larger sizes
 in~\cite{HausmannEA16},
COOL~2 generally outperforms TATL by a large margin, starting from formulas containing at least five modalities or involving nesting of temporal operators; this trend is confirmed by \autoref{fig:nestedGF} which shows
the stepped execution times for the series $\mathsf{Nest}$ that alternates between being satisfiable and unsatisfiable 

In summary, COOL~2 shows promising performance in comparison to TATL and \FaCT{}, as well as for practical applicability.
On graded formulas without fixpoints, COOL~2 scales much better than \FaCT{} with regard to increasing multiplicities.
In the presence of fixpoints, COOL~2 still scales well and can handle multiplicities that should be sufficient for practical use.
The formula series \(\neg \mathsf{ParityToBuechi}\) appears to show the limits of COOL~2 with the current implementation of graded one-step satisfiability checking.
Nonetheless, our results indicate that COOL~2 is capable of automatically
proving or refuting involved properties of (graded) $\omega$-automata and games in reasonable time.  

\section{Conclusion}

We have described and evaluated the current version COOL~2 of the
\emph{CO}algebraic \emph{O}ntology \emph{L}ogic reasoner
(COOL). Future development will include the implementation of
additional instance logics, such as the probabilistic and graded
$\mu$-calculus with linear inequalities, as well as support for the
full coalgebraic $\mu$-calculus via on-the-fly determinisation of
\emph{unrestricted} B\"uchi automata, using the Safra-Piterman
construction.\medskip

\noindent We would like to thank Frederik Hennig for finding and correcting a slight mistake in the Rabin-type formulas in an earlier version of this paper; the corresponding runtimes reported in the table above as well as the full formulas in the appendix have been updated accordingly.


\bibliographystyle{splncs04}
\bibliography{coalgml}
\newpage
\appendix

\section{Appendix: Additional Details}

\subsection{Details for \autoref{rem:graded}}

The following set of modal tableau rules (one-step rules) has been
claimed to be one-step cutfree
complete~\cite{SchroderPattinson09,SnellEA12}:
\begin{equation*}
  \frac{\langle n_1\rangle p_1\land\dots\land\langle n_u\rangle p_u
    \land\neg\langle m_1\rangle q_1\land\dots\land\neg\langle m_v\rangle q_v}
  {\textstyle\sum_{i=1}^ur_ip_i-\sum_{i=1}^vs_iq_i>0}
\end{equation*}
for $r_i,s_i\in\mathbb{N}\setminus\{0\}$, subject to the side condition
\begin{equation*}
  \sum_{i=1}^ur_i(n_i+1)-\sum_{i=1}^vs_im_i \ge 1.
\end{equation*}
The conclusion of the rule denotes a propositional formula (in
conjunctive normal form) representing the Boolean function obtained by
reading the numerical inequality as a constraint on Boolean variables
$p_i,q_i$, with true interpreted as~$1$ and false as~$0$. In the
relevant instance of the one-step logic, models consist of a set $X$,
a valuation~$\tau$ of the variables $p_i,q_i$ as subsets of $X$, and a
finite multiset~$\mu$ over~$X$. Given these data, the
conclusion~$\phi$ of (an instance of) the rule, a purely propositional
formulas, is interpreted as a subset $\sem{\phi}\tau$ of~$X$, using
the Boolean algebra structure of the powerset. The premiss is
evaluated w.r.t.\ satisfaction by~$\mu$, with the expected clauses for
the propositional operators, and with~$\langle n\rangle a$, for a
variable~$a$, being satisfied if $\mu(\tau(a))>n$,
where~$\mu(A)=\sum_{x\in A}\mu(x)$ for $A\subseteq X$. The premiss is
\emph{satisfiable over~$\tau$} if $\sem{\phi}\tau\neq\emptyset$, and
the conclusion is satisfiable over~$\tau$ if it is satisfied by some
multiset~$\mu$ over~$X$ in the sense just defined.

In the one-step logic, the rules are then applied to conjunctions of
\emph{modal literals}, i.e.\ formulas of the form either
$\langle n\rangle a$ or $\neg\langle n\rangle a$, for~$a$ a variable,
requiring an exact match of the rule premiss with a subset of the
conjuncts (thus incorporating weakening into the rule
application). Such formulas are called \emph{one-step clauses}. The
rules are read as tableau rules, i.e.\ to establish that the premiss
is satisfiable, one needs to establish that the conclusions of all
matching rule applications are satisfiable.

The rules are easily seen to be \emph{one-step sound}, i.e.\ if for
any rule instance matching a given one-step clause, the conclusion is
propositionally unsatisfiable over a valuation~$\tau$, then the
premiss is also unsatisfiable over~$\tau$. The rule set is
\emph{one-step complete} if, given a valuation~$\tau$ in the powerset
of~$X$ and a one-step clause~$\chi$, whenever all rule matches
to~$\chi$ have satisfiable conclusions, then~$\chi$ is satisfiable.

We show that the latter property fails. To this end, consider the set
$X=\{a,b,c,d\}$, propositional variables $p_A$ for all $A\subseteq X$,
the valuation~$\tau$ given by $\tau(p_A)=A$, and the one-step
clause~$\chi$ consisting of the positive literals
$\langle 2\rangle p_A$ for all $A\subseteq X$ such that $|A|=2$, and
the negative literal $\neg\langle 6\rangle p_X$. First, note
that~$\chi$ is clearly unsatisfiable over~$\tau$: A multiset~$\mu$
over~$X$ would have to satisfy $\mu(A)\ge 3$ for all $A\subseteq X$
such that $|A|=2$, so at least three of the four elements of~$X$ need
to have multiplicity at least~$2$ under~$\mu$; moreover, if any
element has multiplicity~$0$, then all others need to have
multiplicity at least~$3$. Consequently, the total weight~$\mu(X)$ is
at least~$7$, so~$\mu$ does not satisfy the negative literal
$\neg\langle 6\rangle p_X$. On the other hand, none of the rule
instances matching~$\chi$ have unsatisfiable conclusions. The easiest
way to see this is to note that the rules are, with fairly evident
adaptations, still sound for a real-valued relaxation of the logic
where~$\mu$ may assume non-negative real values (of course,
$\langle n\rangle a$ then means that $\mu(\tau(a))\ge n+1$); under
this semantics, however, $\chi$ is satisfiable over~$\tau$ by taking
$\mu(x)=\frac{3}{2}$ for all $x\in X$.

\subsection{Additional Details for \autoref{sec:mu}}

We give additional details on syntactic notions for the coalgebraic 
$\mu$-calculus, and on predicate liftings. Furthermore, we sketch the
satisfiability checking algorithm, first introduced in~\cite{HausmannSchroder19};
here we give a presentation of the algorithm in terms of automata
and games, tailored towards our implementation in COOL~2.\medskip

Fixpoint operators
\emph{bind} their variable, yielding notions of bound and free
fixpoint variables; a formula then is \emph{closed} if it does not
contain any free variables. A formula is \emph{clean} if every
fixpoint variable is bound at most once in it. We restrict attention
to closed and clean formulas (being aware of issues with formula size
measures~\cite{KupkeEA22b}). Variables $X$ that are bound by 
$\mu X$ then are $\mu$-variables, and variables bound by
$\nu X$ are $\nu$-variables. A variable~$X$ is
\emph{active} in a formula~$\psi$ if~$X$ has a free occurrence in the formula
that is obtained from~$\psi$ by exhaustively
replacing free occurrences of fixpoint variables by their binding
fixpoint formulas.  

E.g. for the formula $\target=\mu X.\,(p\vee\Diamond X)$ we have $\theta(X)=\target$, and the closure
of $\target$ is a graph with nodes 
$\target,p\vee\Diamond\target,p,\Diamond\target$ and edges $\target\to p\vee\Diamond\target$,
$p\vee\Diamond\target\to p$, $p\vee\Diamond\target\to\Diamond\target$, and $\Diamond\target\to\target$.
When we refer to the closure, we typically mean just the set of nodes of this closure graph.\medskip

We give more details on Example~\ref{ex:logics}:\medskip

A coalgebraic modelling of the graded $\mu$-calculus~\cite{DAgostinoVisser02} is obtained by using the functor $F=\mathcal{G}$
that maps a set $C$ to the set $\mathcal{G}C=\{\theta:C\to\mathbb{N}\mid \theta\text{ has finite support}\}$ of finite multisets over~$C$, the similarity type $\Lambda=\{\langle n \rangle,[n]\mid n\in\mathbb{N}\}$, and predicate liftings
\begin{small}
\begin{align*}
\sem{\langle n \rangle}_C(B) &= \{\theta\in \mathcal{G}C\mid \Sigma_{v\in B} \theta(v)>n\} &
\sem{\,[n]\,}_C(B) &= \{\theta\in \mathcal{G}C\mid\Sigma_{v\notin B}\theta(v)\leq n\}
\end{align*}
\end{small}

The \emph{concurrent game
functor} $\mathcal{F}$ maps a set~$C$ to the set  
$\mathcal{F}C=\{(S_1,\ldots,S_n,f)\mid \emptyset\neq S_i,f:\prod\nolimits_{i\leq n}S_i\to C\}$, where the~$S_i$ are viewed as sets of available moves for agent~$i$,
and~$f$ as an outcome function  that evaluates joint
moves of all agents. $\mathcal{F}$-Coalgebras are concurrent game frames~\cite{Pauly02}.
Coalitions $D$ induce sets $S_D=\prod_{i\in D}S_i$ and $S_{\overline{D}}=\prod_{i\in N\setminus D}S_i$, and given $s_D\in S_D$ and $s_{\overline{D}}\in S_{\overline{D}}$, the pair
$(s_D,{s_{\overline{D}}})$ represents an element of $\prod_{i\leq n} S_i$. We use the modal similarity
type $\Lambda=\{\langle D\rangle,[D]\mid D\subseteq N\}$ and the predicate liftings
\begin{align*}
\sem{\langle D\rangle}_C(B)&=\{(S_1,\ldots,S_n)\in \mathcal{F}C\mid \exists s_D\in S_D.\,\forall
s_{\overline{D}}\in S_{\overline{D}}.\, f(s_D,s_{\overline{D}})\in B\}\\
\sem{[D]}_C(B)&=\{(S_1,\ldots,S_n)\in \mathcal{F}C\mid \forall s_D\in S_D.\,\exists
s_{\overline{D}}\in S_{\overline{D}}.\, f(s_D,s_{\overline{D}})\in B\}
\end{align*}
for $B\subseteq C$ and $D\subseteq N$.\medskip

Additional details regarding the satisfiability checking algorithm sketched in the main
paper are as follows.\medskip

Depending on the syntactic structure of the input formula $\target$, it may be possible to
employ simpler determinisation procedures for the
construction of the automaton $\mathsf{B}_\target$, resulting in asymptotically smaller games. 
Currently known bounds for particular fragments of the
coalgebraic $\mu$-calculus are summarised in the following table, where $n$ denotes
the closure size of the target formula, $k$ denotes its alternation-depth, and
where NCBA and LDBA stand for \emph{nondeterministic co-B\"uchi automaton} and 
\emph{limit-deterministic B\"uchi automaton}, respectively.
\begin{center}
\begin{tabular}{|c|c|c|c|c|}
\hline
fragment & type of $\mathsf{A}_\target$ & determinisation& size & rank\\
\hline
alternation-free & NCBA & Miyano-Hayashi~\cite{MiyanoHayashi1984} & $\mathcal{O}(3^n)$ & $2$\\
aconjunctive & LDBA& permutation method~\cite{EsparzaKRS22,HausmannEA18} & $\mathcal{O}((nk)!)$ & $2nk$\\
unrestricted & NBA & Safra-Piterman~\cite{Piterman07} & $2^{\mathcal{O}((nk)\log n)}$ & $2nk$\\
\hline
\end{tabular}
\end{center}

\noindent We sketch the construction of the automata $\mathsf{A}_\target$ and
$\mathsf{B}_\target$ to describe the coalgebraic satisfiability
game.
Recall that the accepted language $L(\mathsf{A})$ of  a parity automaton
$\mathsf{A}=(Q,\Sigma,\delta,v_0,\Omega)$ with priority function $\Omega:Q\to\mathbb{N}$
consists of those infinite words over $\Sigma$
on which there is an accepting run of $\mathsf{A}$, where a run is accepting
if and only if the maximal priority (according to $\Omega$) that is visited infinitely often is even.
The logical connectives are captured by the alphabet of the
automaton. Propositional connectives are treated by letters from a set $\Sigma_p$,
also encoding choices of disjuncts using letters of the shape $(\psi_1\vee\psi_2,b)$ where
$\psi_1\vee\psi_2\in\mathsf{cl}$ and $b\in\{1,2\}$. For modalities, the automaton needs
to work on the conjunctive satisfiability of a set of operators, so
we put
\begin{align*}
\Sigma_s:=\{\kappa\in\Pow(\mathsf{cl})\mid \forall \psi\in\kappa.\,\exists\hearts\in\Lambda.\,\psi=\hearts\phi\}.
\end{align*}
The alphabet of the automata is
$\Sigma=\Sigma_p\cup\Sigma_s$. Furthermore,
we let $\mathsf{choices}=\{w\in\Sigma_p^*\mid |w|\leq n^2\}$ denote
the set of propositional words of length at most $n^2$.

The nondeterministic parity automaton 
$\mathsf{A}_\target=(\mathsf{cl},\Sigma,\Delta,\target,\Omega')$ reads
sequences $w\in\Sigma^\omega$ of formula manipulations and traces formulas through the
closure. For the transition function we have, for example
$\Delta(\psi_1\vee\psi_2,(\psi_1\vee\psi_2,b))=\{\psi_b\}$ for $b\in\{1,2\}$,
$\Delta(\psi_1\wedge\psi_1,(\psi_1\wedge\psi_1))=\{\psi_1,\psi_2\}$,
$\Delta(\hearts\psi,\kappa)=\{\psi\}$ if $\hearts\psi\in\kappa\in\Sigma_s$
and $\Delta(\hearts\psi,\kappa)=\emptyset$ otherwise.
The priority function $\Omega'$ is defined by the alternation of
least and greatest fixpoints in the formula, and 
$\mathsf{A}_\target$ accepts words encoding infinite traversals
through formulas
where the outermost unfolded fixpoint is a least fixpoint.
We now consider the deterministic parity automaton $\mathsf{B}_\target=(\detcarrier,\Sigma,\delta,v_0,\Omega)$ 
that is obtained from $\mathsf{A}_\target$
by co-determinisation ($L(\mathsf{A}_\target)=\overline{L(\mathsf{B}_\target})$), noting that
$\mathsf{B}_\target$ is a parity automaton with at most $\prios$ priorities by the results from~\cite{Piterman07}.
Then $\mathsf{B}_\target$ accepts infinite traversals through
formulas for which there is no formulas trace (run of $\mathsf{A}_\target$) on which the
outermost fixpoint that
is unfolded infinitely often is a least fixpoint.

The set $\detcarrier$ consists of macro-states, that is, data
structures organising elements of $\mathsf{cl}$ in way that depends on
the concrete determinisation construction that is used (see the table
above), e.g., as Safra-trees~\cite{Piterman07} or
permutations~\cite{EsparzaKRS22}.  The \emph{labelling} function
$l\colon\detcarrier\to\Pow(\mathsf{cl})$ assigns to each macro-state
$v\in\detcarrier$ its \emph{label} $l(v)$, i.e.\ the set of formulas
that occur in~$v$.  We denote by $\mathsf{cores}$ and
$\mathsf{states}$ the sets of all states in $\detcarrier$ that have an
unsaturated or a saturated label, respectively; here, a set of
formulas is \emph{saturated} if all its elements are either modalised
formulas $\hearts\psi$ or $\top$, and unsaturated otherwise.  We
extend $\delta$ to words over $\Sigma$ in the usual way and to subsets
of $\Sigma_s$ by putting $\delta(v,A)=\{\delta(v,a)\mid a\in A\}$ for
$v\in\detcarrier$, $A\subseteq\Sigma_s$.

\begin{defn}[One-step satisfiability
  problem~\cite{Schroder07,SchroderPattinson08,MyersEA09}]~\label{defn:onestep}
  Let $V$ be a finite set, $\Lambda(V) = \lbrace \heartsuit a \mid a
  \in V, \heartsuit \in \Lambda \rbrace$, and let $\osZero\subseteq\Pow(V)$. We interpret
  $a\in V$ and $\osOne\subseteq \Lambda(V)$ over~$\osZero$ by
  \[
    \sem{a}^\osZero_0  = \{u\in \osZero\mid a\in u\} \qquad
    \sem{\osOne}^\osZero_1
                 =\textstyle\bigcap_{\hearts
                 a\in
                 \osOne}\sem{\hearts}_\osZero\sem{a}^\osZero_0.
  \]
  We refer to the data $(\osOne,\osZero)$ as a \emph{one-step pair}
  (over~$V$) 
  and say that $(\osOne,\osZero)$ is \emph{satisfiable} (over~$F$) if
  $\sem{\osOne}^\osZero_1\neq\emptyset$.
\end{defn}
  
\begin{expl}\label{ex:onestepsat}
  \begin{enumerate}[wide]
  \item For the \emph{relational modal $\mu$-calculus}
    (\autoref{ex:logics}.1.), where $\Lambda=\{\Diamond,\Box\}$,
    the one-step satisfiability problem is to decide, for a given
    one-step pair $(\osOne,\osZero)$ over~$V$, whether there is
    $A\in \sem{\osOne}^\osZero_1$, that is, a subset $A\in\Pow \osZero$ such that for
    each $\Diamond a\in \osOne$, there is $u\in A$ such that $a\in u$, and
    for each $\Box b\in \osOne$ and each $u\in A$, $b\in u$.  Equivalently,
    one needs to check that for each~$\Diamond a\in \osOne$ there is
    $u\in \osZero$ such that $a\in u$ and moreover $b\in u$ for all
    $\Box b\in \osOne$. 
  \item For the \emph{graded $\mu$-calculus}
    (\autoref{ex:logics}.2.), the one-step
    satisfiability problem is to decide, for a one-step
    pair~$(\osOne,\osZero)$, whether there is a multiset
    $\beta\in\Bag \osZero$ such that
    $\sum_{u\in \osZero\mid a\in u}\beta(u)>m$ for each
    $\langle m\rangle a\in \osOne$ and
    $\sum_{u\in \osZero\mid a\notin u}\beta(u)\leq m$ for each
    $[m]a\in \osOne$. This problem can be solved via a nondeterministic
    algorithm that goes through all $u\in \osZero$,
    guessing multiplicities $\beta(u)\in\{0,\dots,m+1\}$ where~$m$ is
    the greatest index of any diamond modality $\langle m\rangle$ that
    occurs in~$\osOne$. This multiplicity is used to update~$|V|$
    counters that keep track of the total
    measure~$\beta(\Sem{a}^\osZero_0)$ for $a\in V$ and then
    forgotten. Once all multiplicities have been guessed, the
    algorithm verifies that $\beta\in\Sem{\osOne}^\osZero_1$, using
    only the final counter
    values~\cite[Lemma~1]{KupfermanEA02}.
  \end{enumerate}
\end{expl}

Recall (e.g.\cite{Graedel02}) that \emph{parity games} are
history-free determined infinite-duration two-player games given by
data $G=(V_\exists,V_\forall,E,v_0,\Omega)$, consisting of disjoint
sets $V_\exists$ and $V_\forall$ of nodes belonging to the existential
and universal player, respectively, a set $E\subseteq V\times V$ of
edges (where $V=V_\exists\cup V_\forall$), an initial node~$v_0$, and
a priority function $\Omega:E\to\mathbb{N}$ that assigns priorities
$\Omega(e)$ to \emph{edges} $e\in E$.  A \emph{play} is a finite or infinite
sequence $\pi=v_0,v_1,\ldots$ of nodes such that $(v_i,v_{i+1})\in E$
whenever applicable.  Finite plays are required to end in nodes
without outgoing moves, and then are won by the player that
does not own the last node in the play; infinite plays
$\pi\in V^\omega$ are won by the existential player if and only if the
maximal priority visited in $\pi$ is even (and by the universal player
otherwise).  The existential player wins the game $G$ if she has a
strategy to move at her nodes such that she wins every play that is
compatible with this strategy; otherwise, the universal player wins
$G$.

Now we are ready to characterise satisfiability in the coalgebraic $\mu$-calculus
by parity games. Recall that $\mathsf{B}_\target=(\detcarrier,\Sigma,\delta,v_0,\Omega)$
is a deterministic parity automaton with $\prios$ priorities.

\begin{defn}[Satisfiability games]\label{defn:satgames}
The \emph{satisfiability game} $G_\target=(V_\exists,V_\forall,E,v_0,\Omega)$ for $\target$ is a parity game
with sets of nodes $V_\exists=\detcarrier$ and $V_\forall=\detcarrier\times\Pow(\Sigma_s)$.
The other components of the game are defined by the following table, where $\Omega(v,\tau)$ denotes the maximal $\Omega$-priority that is visited by the partial run
of $\mathsf{B}_\target$ that leads from $v$ to $\delta(v,\tau)$ by reading $\tau$ letter by letter.
\begin{center}
\begin{tabular}{|c|c|c|c|}
\hline
node & owner & moves to & priority\\
\hline
$v\in\mathsf{cores}$ & $\exists$ &  
$\{\delta(v,\tau)\in \mathsf{states}\mid \tau\in\mathsf{choices}\}$ &
$\Omega(v,\tau)$\\
$v\in\mathsf{states}$ & $\exists$ &  
$\{(v,\Xi)\in V_\forall\mid \sem{l(v)}^{l[\delta(v,\Xi)]}_1\neq\emptyset\}$ & $\Omega(v)$\\
$(v,\Xi)\in V_\forall$ & $\forall$ & $\{\delta(v,\kappa)\mid \kappa\in\Xi\}$ & $0$\\
\hline
\end{tabular}%
\end{center}
\end{defn}
\noindent  Thus the existential player
attempts to show the existence of a specific sub-automaton of
$\mathsf{B}_\target$, intuitively by selecting, at each core node~$v$
of~$\mathsf{B}_\target$, some non-modal word $\tau$ that saturates $v$, leading to a state node. 
For modal steps at state nodes~$v$, the existential player has to provide a set 
$\Xi$ of letters (selecting a set of modal out-edges of~$v$) such that the label of $v$ is
one-step satisfiable in the labels of the selected successors of $v$ in $\mathsf{B}_\target$.
The universal player then can challenge any letter 
$\kappa\in\Xi$ and perform the corresponding modal step by moving to $\delta(v,\kappa)$.
The existential player has to make her choices in such a way that for every compound
word that results
from the choices, the corresponding run of $\mathsf{B}_\target$ is accepting 
(that is, does not contain  infinite deferrals of
least fixpoints), and never visits nodes with $\bot$ in the label. 

Given a set $G\subseteq \detcarrier$, we let $\mathsf{win}^\exists_G$
and $\mathsf{win}^\forall_G$ denote the winning regions for the
existential and the universal player, respectively, in the partial
game $G_\target|_G$ that is obtained from $G_\target$ by removing all
nodes that are not contained in $G$.  Nodes in this partial game
$G_\target|_G$ may be undetermined so that we in general do not have
$\mathsf{win}^\exists_G\cup\mathsf{win}^\forall_G=G$. However, we do
have
$\mathsf{win}^\exists_{\detcarrier}\cup
\mathsf{win}^\forall_{\detcarrier}={\detcarrier}$.  While the number
of nodes in $G_\target|_G$ is doubly exponential in $n$ ($\Sigma_s$ is
singly exponential, so $\Pow(\Sigma_s)$ is doubly
exponential),
we use a fixpoint computation over (subsets of) $\detcarrier$~\cite{HausmannSchroder19}
to compute the sets $\mathsf{win}^\exists_G$ and $\mathsf{win}^\forall_G$ in singly exponential time.

\begin{defn}[Game solving]\label{def:propagation} 
Given a set $G\subseteq \detcarrier$, we define the \emph{one-step solving function}
$f_G:\Pow(G)^{\prios}\to \Pow(G)$ by
  \begin{align*}
  f_G(\mathbf{X})=&
  \{v\in \mathsf{cores}\mid \exists \tau\in\mathsf{choices}.\,
                  \delta(v,\tau)\in X_{\detprio(v,\tau)}\cap\mathsf{states}\}\,\cup\\
 & \{v\in \mathsf{states} \mid \sem{l(v)}^{l[\delta(v,\Sigma_s)\cap X_{\detprio(v)}]}_1\neq\emptyset\}
 \end{align*}
where $\mathbf{X}=X_1,\ldots, X_{\prios} \subseteq G$.
This function intuitively encodes
one step in the partial game $G_\target|_G$ that is obtained from $G_\target$ by removing all nodes
that are not contained in $G$. The winning regions in $G_\target|_G$
then can be characterised by the nested fixpoints $\mathbf{E}_G$ and $\mathbf{A}_G$, defined by
\begin{align*}
\mathbf{E}_G&=\eta_{\prios} X_{\prios}.\,\ldots \eta_1 X_1. f_G(\mathbf{X}) &
\mathbf{A}_G&=\overline{\eta_{\prios}} X_{\prios}\,\ldots \overline{\eta_1} X_1. \overline{f_G}(\overline{\mathbf{X}}),
\end{align*}
where $\eta_i= \mu$ for odd $i$,
$\eta_i=\nu$ for even~$i$, where $\overline{\nu}=\mu$ and
$\overline{\mu}=\nu$, and where $\overline{f_G}(\overline{\mathbf{X}})=\detcarrier\setminus
f_G({\mathbf{X}})$.
\end{defn}
We note that nodes in the partial game $G_\target|_G$ may be undetermined so that we in general do not have
$\mathbf{E}_G\cup\mathbf{A}_G=G$. However, we do have $\mathbf{E}_{\detcarrier}\cup
\mathbf{A}_{\detcarrier}={\detcarrier}$.

\begin{algorithm}\label{alg:global}
  Initialise set of \emph{unexpanded} nodes $U=\{v_0\}$ and \emph{expanded} nodes $G=\emptyset$.
\begin{enumerate}
\item Expansion: Choose some unexpanded node $u\in U$, remove~$u$ from
  $U$, and add~$u$ to $G$.  Add to $U$ all nodes in the sets
  $\{\delta(u,\tau)\in\mathsf{states}\mid\tau\in\mathsf{choices}\}\setminus G$ 
  (if $u\in\mathsf{cores}$) or
  $\{\delta(u,\kappa)\mid \kappa\in\Sigma_s,\kappa\subseteq
  l(u)\}\setminus G$ (if $u\in\mathsf{states}$).
\item Optional game solving: Compute $\mathsf{win}^\exists_G$ and/or $\mathsf{win}^\forall_G$. If
$v_0\in\mathsf{win}^\exists_G$, then return `satisfiable`, if $v_0\in\mathsf{win}^\forall_G$,
then return `unsatisfiable`.
\item If $U\neq\emptyset$, then continue with Step~1. Otherwise, $G=\detcarrier$; continue
with Step~4.
\item Final game solving: Compute $\mathsf{win}^\exists_G$. If
$v_0\in\mathsf{win}^\exists_G$, then return `satisfiable`, otherwise return
`unsatisfiable`.
\end{enumerate}
\caption{Satisfiability checking by global caching; input: formula~$\target$;}
\end{algorithm}

\begin{theorem}[\hspace{-3pt}~\cite{HausmannSchroder19}]
  Existential player wins $G_\target$ if and only if $\target$ is
  satisfiable, and Algorithm~1 computes the winning regions
  in~$G_\target$ on-the-fly.
\end{theorem}
(The proof given in~\cite{HausmannSchroder19} elides the game
formulation and instead shows directly that the fixpoint captures
satisfiability; however, equivalence of the game and the fixpoint
computation is straightforward.)

\subsection{Additional Details for \autoref{sec:eval}}
The omitted formulas for the parity conditions in \autoref{sec:eval} are defined as follows.
{\small
\begin{align*}
\mathsf{Parity}(n,k)&=\mu X_k.\, \nu X_{k-1}.\,\ldots. \nu X_2.\,\mu X_1.\, \bigvee_{1\leq i\leq k} p_i\wedge\langle n\rangle X_i 
\end{align*}
}%
These formulas express that
\emph{`there is an $n+1$-branching tree starting here in which each path satisfies the
parity condition encoded by the priorities $p_i$'}.
This property implies that
\emph{`there is an $n+1$-branching tree such that for every path 
there is an even priority $p_i$ that occurs infinitely often on the path
and priorities larger than $p_i$ occur only finitely often on the path'}:
{\small
\begin{align*}
\mathsf{Buechi}(n,k) &=
\mu X.\, \langle n\rangle X\vee \bigvee_{1\leq i\leq k, i\text{ even}}
\nu Y.\, \mu Z.\, \bigwedge_{j>i}\neg p_j\wedge ((p_i\wedge \langle n\rangle Y) \vee \langle n\rangle Z))
\end{align*}
}%

The formulas  $\mathsf{Buechi}(f,\psi)$,  $\mathsf{RabinPair}(i,f,\psi)$ and  $\mathsf{Rabin}(k,\psi)$ for the rabin properties
$\langle i_j,f_j\rangle_{j\leq k}$ in \autoref{sec:eval} are defined as follows.
 {\small
  \begin{align*}
 \mathsf{Buechi}(f,\psi)&=\nu X.\,\mu Y.\,((f \wedge \psi(X))\vee (\neg f\wedge\psi(Y))) \\
 \mathsf{RabinPair}(i,f,\psi)&=\mu X.\,\nu Y.\,\mu Z.\,(f\wedge\psi(X))\vee 
 (\neg f\wedge i \wedge\psi(Y))\vee (\neg f\wedge\neg i\wedge\psi(Z)) \\
 \mathsf{Rabin}(k,\psi)&=\mu X_{2k+1}.\,\mathsf{Disj}(2k,[],\psi)\\
 \mathsf{Disj}(c,p,\psi)&=\bigvee_{j\notin p}\nu X^{p:j}_{c}.\,\mu X^{p:j}_{c-1}.\,
 \mathsf{Disj}(c-2,p:j,\psi)\\
 \mathsf{Disj}(0,p,\psi)&=(f_{p(1)} \wedge \psi(X_{2k+1}))\vee\\
 & \quad\,\,\bigvee_{j\leq k}  
 ((\bigwedge_{j'\leq j}\neg f_{p(j')}\wedge i_{p(j)}\wedge \psi(X^{p|j}_{2(k-j)+2}))\vee\\
&\qquad\quad\,\,\, (\bigwedge_{j'\leq j}\neg f_{p(j')}\wedge \neg i_{p(j)}\wedge\psi(X^{p|j}_{2(k-j)+1})))
 \end{align*}
 }
 Here, $p$ is a partial permutation over $[k]=\{1,2,\ldots,k\}$, that is, $p$ is a list of elements of $[k]$
 without duplicates; the empty permutation is denoted by $[]$. We write $j\notin p$
 to denote the fact that $j\in[k]$ does not occur in the partial permutation
 $p$. By $p:j$ we denote the partial permutation that is obtained by concatenating $p$ and $j$
 (that is, by appending $j$ to the end of the list $p$). Finally, $p|i$ denotes the permutation
 that is obtained from $p$ by keeping just the first $i$ entries, and $p(j)$ denotes the $j$-th entry
 in $p$.\medskip

 The full formulas for B\"uchi and Rabin games mentioned in \autoref{sec:eval} are as follows.\medskip
 
 The formula $\theta =\mathsf{AG}((v_\exists\wedge \neg v_\forall)
  \vee(\neg v_\exists\wedge v_\forall))$
 expresses that every node belongs to exactly one of the two players $v_\exists$ or $v_\forall$. We also introduce graded formulas $\mathsf{Cpre}^n_i(X)$ stating that player $i\in\{\exists,\forall\}$ can ensure that $X$ is reached in the next step with
 at least (all but at most) $n$ moves:
 {\small
\begin{align*}
 \mathsf{Cpre}^n_\exists(X) &=(v_\exists \wedge \langle n\rangle X) \vee (v_\forall \wedge [n] X) &
 \mathsf{Cpre}^n_\forall(X) &=(v_\forall \wedge \langle n\rangle X) \vee (v_\exists \wedge [n] X)
 \end{align*}
}%
 The winning regions in graded B\"uchi and Rabin games then are defined by
 {\small
\begin{align*}
 \mathsf{BuechiG}(f,n)&=\theta\land\mathsf{Buechi}(f,\mathsf{Cpre}^n_\exists) &
  \mathsf{RabinG}(k,n)&= \theta\land\mathsf{Rabin}(k, \mathsf{Cpre}^n_\exists)
 \end{align*}
 }%
 
The comparison between TATL and COOL~2 is based on the following formulas taken from~\cite{David13}:

\begin{footnotesize}
\begin{enumerate}
\item $p$ 
\item $p\land q$ 
\item $p\lor q$ 
\item $p\to q$ 
\item \(\ATLquant{1}X p\) 
\item$\ATLquant{1}F p$ 
\item $\ATLquant{1}G p$ 
\item $\ATLquant{1}p U q$ 
\item $\neg\ATLquant{1}p U q$ 
\item $\neg\ATLquant{1}F p$ 
\item $\ATLquant{1,2}p U q \wedge \ATLquant{1,2}X r$ 
\item $\ATLquant{1,2}p U q \wedge \ATLquant{3,4}X r$ 
\item $\ATLquant{1,2}p U q \wedge \ATLquant{2,3}X r$ 
\item $\ATLquant{2,1}p U q \wedge \ATLquant{3,2}X r$ 
\item $\ATLquant{}p U q \wedge \ATLquant{1,2}X r$ 
\item $\neg\ATLquant{1,2}X p \wedge \ATLquant{1}G p$ 
\item $\neg\ATLquant{1,2}X p \wedge \ATLquant{1,2,3}G p$ 
\item $\neg p \vee \ATLquant{1}F p$ 
\item $p \wedge \neg p $
\item $(p \wedge q) \wedge \ATLquant{1}G \neg(p \wedge q)$ 
\item $\ATLquant{1}G p \wedge \neg\ATLquant{2}F \ATLquant{1}G p$ 
\item $\ATLquant{1}X p \wedge \neg\ATLquant{1}X p$ 
\item $\ATLquant{1}p U q \vee \neg\ATLquant{1}G q$
\item $\ATLquant{1,2}p U (\neg\ATLquant{1}G p)$
\item $\ATLquant{1}(\neg\ATLquant{1,2}G p) U q$
\item $\ATLquant{}G\ATLquant{}p U q$ 
\item $\neg\ATLquant{1}G p \wedge \ATLquant{1,2}X p \wedge \neg\ATLquant{2}X \neg p$ 
\item $\ATLquant{1}X p \wedge \ATLquant{2}X q \wedge \ATLquant{1,2}X r \wedge \neg\ATLquant{1}X r \wedge \neg\ATLquant{3}X q$ 
\item $\neg\ATLquant{1}X r \wedge \neg\ATLquant{3}X q \wedge \ATLquant{1}X p \wedge \ATLquant{2}X q \wedge \ATLquant{1,2}X r$ 
\item $\neg\ATLquant{1}X r \wedge \ATLquant{1}X p \wedge \ATLquant{2}X q \wedge \neg\ATLquant{3}X q \wedge \ATLquant{1,2}X r$ 
\item $\ATLquant{1,2,3}G\ATLquant{2,3,4}G( p \wedge q)$ 
\item $\ATLquant{1,2,3}G\ATLquant{2,3}G(p\wedge q) \wedge \ATLquant{4}X \neg p$ 
\item $\neg\neg\ATLquant{1}p U q$ 
\item $\neg(\ATLquant{1}G p \vee \ATLquant{1}G \neg p)$ 
\item $\neg(\ATLquant{1}G p \wedge \ATLquant{1}G \neg p)$ 
\item $\neg\ATLquant{1}p U\neg\ATLquant{2}q U r$ 
\item $\ATLquant{1}G \neg q \wedge\ATLquant{2}p U q$ 
\item $\ATLquant{1}G p \wedge \neg\ATLquant{1,2}G p$ 
\item $\neg\ATLquant{1}X p \wedge\ATLquant{2}X \neg p$ 
\item $\ATLquant{1}X p \wedge \ATLquant{2}X \neg p$ 
\item $\ATLquant{1}p U q \wedge \ATLquant{2}q U r \wedge \ATLquant{2}G \neg r$
\item $\ATLquant{1}p U q \wedge \ATLquant{2}q U r \wedge \ATLquant{1}G \neg r$
\end{enumerate}

\end{footnotesize}

The results of the comparison between COOL~2 and TATL on these formulas are shown
in \autoref{fig:tatl}.

\begin{figure}
  \centering
  \begin{tikzpicture}
    \begin{axis}[
      ybar=0,
      minor x tick num=1,
      xminorgrids,
      minor tick length=0,
      bar width = .25em,
      every axis y label/.style={at={(ticklabel cs:0.5)},rotate=90,anchor=center},
      every axis x label/.style={at={(ticklabel cs:0.5)},anchor=center},
      tiny,
      width=\linewidth,
      height=4cm,
      legend style={at={(0.15,0.95)},anchor=north},
      xlabel={formula},
      ylabel={runtime (ms)},
      ymin=0,
      ymax=1.3,
      xmin=0.5,
      xmax=42.5,
      xtick distance = 1,
      x tick label style={rotate=90},
      x tick style = transparent,
      legend entries={COOL on-the-fly,TATL}]

      \addplot[draw=blue, fill=blue!20!white] table [col sep=comma, x=row, y=mean] {benchmarks/mergedcoolResults.csv};
      \addplot[draw=black, fill=black!20!white] table [col sep=comma, x=row, y=mean] {benchmarks/mergedtatlResults.csv};
    \end{axis}
  \end{tikzpicture}
  \caption{Runtimes on selected ATL formulas~\cite{David13}}\label{fig:tatl}

\end{figure}

Furthermore, we turn formula $36$, that is, the formula $\neg\ATLquant{1} p\, \mathsf{U}\,
\neg\ATLquant{2} q\, \mathsf{U}\,r$ into a formula series with increasing number of nested temporal operators, defined inductively
by 
\begin{align*}
\psi(0)&=\neg\ATLquant{2} q\, \mathsf{U}\,r\\
\psi(i+1)&=\neg\ATLquant{(i\bmod 2)+1} p_{i\bmod 2}\, \mathsf{U}\, \psi(i) \tag{$i>0$}
\end{align*}
The according experiment results depicted in \autoref{fig:nestedU} show that COOL~2 outperforms TATL on this formula series by a large margin,
with runtime for COOL~2 remaining almost constant as $n$ increases while TATL quickly runs into timeouts of $60$ seconds. This presumably is due to the nesting of temporal operators.

\begin{center}
\begin{figure}
\begin{minipage}[t][][t]{.5\linewidth}
  \begin{tikzpicture}
    \begin{semilogyaxis}[
      minor tick num=0,
      xtick distance = 1,
      log basis y = 10,
      log ticks with fixed point,
      every axis y label/.style=
      {at={(ticklabel cs:0.5)},rotate=90,anchor=center},
      every axis x label/.style=
      {at={(ticklabel cs:0.5)},anchor=center},
      tiny,
      width=\linewidth,
      height=4cm,
      legend columns=3,
      legend style={at={(0.5,-0.2)},anchor=north},
      ymode=log,
      xlabel={$n$ (nesting depth)},
      ylabel={runtime (s)},
      xmin=0,
      xmax=10,
      ymin=0,
      ymax=60,
      legend entries={COOL on-the-fly,COOL,TATL}]

      \addplot [color=blue, mark=*] table [col sep=comma, x expr=\coordindex, y=mean] {benchmarks/nestedUCOOL.csv};
      \addplot [color=red, mark=square*] table [col sep=comma, x expr=\coordindex, y=mean] {benchmarks/nestedUCOOLOnce.csv};
      \addplot [color=black, mark=triangle*] table [col sep=comma, x expr=\coordindex, y=mean] {benchmarks/nestedUTATL.csv};
    \end{semilogyaxis}
  \end{tikzpicture}
  \caption{Runtimes for $\psi(n)$}\label{fig:nestedU}
\end{minipage}
\end{figure}
\end{center}

\end{document}
